\documentstyle[preprint,revtex]{aps}
\begin{document}
\draft
\begin{title}
ON THE ORIGIN OF THE --4.4 eV BAND IN CdTe(100)
\end{title}
\author{D. Olgu\'{\i}n. and R. Baquero.}

\begin{instit}
Departamento de F\'{\i}sica CINVESTAV-IPN
A.P. 14-740, 07000 M\'exico, D.F.
\end{instit}
\begin{abstract}
We calculate the bulk- (infinite system), (100)-bulk-projected- and
(100)-Surface-projected Green's functions using the Surface Green's Function
Matching method (SGFM) and an empirical tight-binding hamiltonian  with
tight-binding parameters (TBP) that
describe well the bulk band structure of CdTe. In particular, we
analyze
the band (B--4) arising at --4.4 eV from the top of the valence band
at $\Gamma$
according to the results of Niles and H\"ochst and at -4.6 eV according
to Gawlik {\it et al.} both obtained by
Angle-resolved photoelectron spectroscopy (ARPES). We give the first
theoretical
description of this band.
\end{abstract}
\narrowtext
\end{document}